\documentclass[11pt,dvips]{article}
\usepackage{graphicx}
\usepackage{amsmath}
\usepackage{amssymb}
\usepackage{latexsym}
\begin{document}
\thispagestyle{empty}
\begin{center}

\vspace{1.8cm}

%%%%%%%%%%%%%%%%%%%%%%%%%%%%%%%%%%%%%%%%%%%%%%%%%%%%%%%%%%%%%%%%%%%%%%%%%%%%%%%%%%%%%%%%
{\bf Three dimensional quantum key distribution in the presence of several eavesdroppers}\\
%%%%%%%%%%%%%%%%%%%%%%%%%%%%%%%%%%%%%%%%%%%%%%%%%%%%%%%%%%%%%%%%%%%%%%%%%%%%%%%%%%%%%%%

\vspace{1.5cm}

{\bf M. Daoud}$^{a,b}${\footnote { email: daoud@pks.mpg.de,
m$_-$daoud@hotmail.com}} and {\bf H. Ez-zahraouy}$^c${\footnote {
email: ezahamid@fsr.ac.ma}}

\vspace{0.5cm}

$^a$ {\it Max Planck Institute for the Physics of Complex Systems \\
 Dresden,
Germany}\\ \vspace{0.2cm}
$^b$ {\it Department of Physics, Faculty of Sciences, University Ibn Zohr,\\
 Agadir ,
Morocco}\\ \vspace{0.2cm} $^c$ {\it
LMPHE (URAC),  Faculty of Sciences, University Mohammed V-Agdal,\\
 Rabat,
Morocco}\\[1em]

\vspace{3cm}
{\bf Abstract}
\end{center}

\baselineskip=18pt
\medskip

Quantum key distribution based on encoding in three dimensional
systems in the presence of several eavesdroppers is proposed. This
extends the BB84 protocol in the presence of many eavesdroppers
where two-level quantum systems (qubits) are replaced by three-level
systems (qutrits). We discuss the scenarios involving  two, three
and four complementary bases. We derive the explicit form of Alice
and Bob mutual information and the information gained by each
eavesdropper. In particular, we show that, in the presence of only
one eavesdropper,
 the protocol involving four bases is safer than the other ones. However,
for two eavesdroppers,  the security is strongly dependent on the
attack probabilities. The effect of a large number of eavesdroppers
is also investigated.

\newpage
\section{Introduction}

 Inspired by Wiesner's ideas \cite{Wiesner}, Bennett and Brassard  proposed in 1984 ~\cite{BB84}
 a new approach to
cryptography by developing a key distribution protocol , now known
as BB84. Since then Quantum Key Distribution (QKD) constitutes one
the most investigated concept in the quantum information theory. QKD
provides a scheme to ensure a secure communication between two
legitimate parties (usually called Alice and Bob) using quantum
states that belong to non compatible bases. Quantum mechanics
ensures that in quantum cryptographic protocols, the presence of an
eavesdropper (often called Eve) in the communication channel can be
detected through  disturbances to the transmitted message. In the
BB84 protocol~\cite{BB84}, Alice and Bob randomly choose between two
complementary bases and the information of each basis is encoded
using the orthogonal states of a two dimensional quantum system.
This protocol was extended in different ways. The first extension
was proposed in~\cite{Bruss PRL 98}-\cite{Bechmann1} by adding an
extra basis that is mutually unbiased compared to the other two in a
two dimensional system. Cryptographic schemes, extending the BB84
model, based on $d$ level quantum systems with $M$ mutually unbiased
bases were also developed. In this sense, a protocol using $d = 4$
states and $M = 2$ bases was studied in~\cite{Bechmann2} and the
case of $d=3$ states and $M = 4$ bases was presented
in~\cite{Bechmann3}. This provided a way to formulate a generalized
quantum key distribution involving quantum systems with arbitrary
dimension $d$ and using $M$ mutually complementary bases
\cite{Bourennane}-\cite{Cerf}.

The main task of quantum protocols, mentioned above as well as many
others proposed in the literature, is traditionally to provide
secure communications against only one eavesdropper. Recently, in
\cite{hamid}, the authors discussed a scenario involving several
eavesdroppers. This extends the BB84 protocol to take into account
the effects induced by many potential eavesdroppers. In this paper,
we replace, in the model introduced in \cite{hamid}, the two-level
quantum systems (qubits) by  optical biphoton qutrits. At this
stage, it is important to note that three level optical systems
constitute promising objects of modern quantum information and
quantum cryptography. Indeed, the realization of optical qutrits
with light has been approached using the polarization states of two
indistinguishable photons -a biphoton \cite{Bogdanov} and their
experimental manipulation was discussed in \cite{Lanyon}. Also,
quantum encoding based on polarization states of a biphoton was
examined in \cite{Bregman}.

To investigate quantum key distribution based on a three level
optical system, we shall first develop, in the second section, a
method to construct the phase states of a biphoton-system and we
define the discrete operations generating  four mutually unbiased
bases from which Alice can choose to encode her message. A second
facet of this work concerns the mutual information between Alice and
Bob and the information intercepted by the eavesdroppers which
employ the intercept-resend strategy. The explicit expressions of
mutual informations as well as quantum errors are given, in the
third section, in terms of the number $N$ of eavesdroppers, the
attack probabilities and the number $M$ of complementary bases used
by the sender to encode her message. In the last section, analysis
of the security of the model in the particular cases of $N=1$ and
$N=2$ are presented. We also discuss the case where the $N$
eavesdroppers are collaborating (all intercept the sent message with
identical probabilities). Concluding remarks close this paper.

%%%%%%%%%%%%%%%%%%%%%%%%%%%%%%%%%%%%%%%%%%%%%%%%%%%%%%%%%%%%%%%%%%%%%%%%
 \section{Qutrits, Phase states and mutually unbiased bases}
 %%%%%%%%%%%%%%%%%%%%%%%%%%%%%%%%%%%%%%%%%%%%%%%%%%%%%%%%%%%%%%%%%%%%%%%

%%%%%%%%%%%%%%%%%%%%%%%%%%%%%%%%%%%%%%%%%%%%%%%%%%%%%%%%%%%%%%%%%%%%%%%%
 \subsection{Phase states }
 %%%%%%%%%%%%%%%%%%%%%%%%%%%%%%%%%%%%%%%%%%%%%%%%%%%%%%%%%%%%%%%%%%%%%%%

The biphoton qutrits are considered as superpositions of the three
dimensional Fock space corresponding to the three possibilities of
distributing two indistinguishable photons in two polarization modes
horizontal $(h)$ and vertical $(v)$. The Fock space of purely
polarized biphoton states is defined by
\begin{equation}
{\cal F} = \{ \vert n_{h} , n_{v} \rangle ~ , ~  n_{h} + n_{v} = 2
\}
\end{equation}
where $\vert n_{h} , n_{v} \rangle$ is a Fock representation of
$n_h$ ($n_v$) horizontally (vertically) polarized photons. They are
given by
\begin{equation}
\vert n_h , n_v \rangle =
\frac{(a^+_h)^{n_h}}{\sqrt {n_h!}}\frac{(a^+_h)^{n_v}}{\sqrt{ n_v!}}\vert 0
0\rangle.
\end{equation}
The vector $\vert 0 , 0 \rangle$  is the vacuum state and the objects $a^+_h$ and
$a^+_v$ are the creation operators of photons with horizontal and
vertical polarizations (with given equal frequencies and given
identical propagation directions). The annihilation operators are
defined as usual $( a_h^- = (a_h^+)^{\dagger} ,  a_v^- = (a_v^+)^{\dagger} )$.

To introduce the phase states, we first define the unitary phase
operator as in \cite{Pegg}
\begin{equation}
E = \vert 2 , 0 \rangle \langle 0 , 2 \vert + \vert 1 , 1 \rangle
\langle 2 , 0 \vert + \vert 0, 2 \rangle \langle 1 , 1 \vert.
\end{equation}
It is unitary. To find the phase states corresponding to this  three
level system, let us consider the eigenvalue equation
\begin{equation}
E \vert z \rangle = z \vert z \rangle, \qquad z \in {\bf C}.
\end{equation}
By expanding the state $ \vert z \rangle$ as a linear combination of
the vector states of ${\cal F}$, it is easy to see that the
eigenvalue $z$ is given by
\begin{equation}
 z = q^m = \exp\bigg(i\frac{2\pi m}{3} \bigg)  ~ {\rm with} ~ m = 0, 1,
 2,
\end{equation}
and the normalized eigenstates of the operator $E$ (the phase
states) rewrite
\begin{equation}
\vert m \rangle = \frac{1}{\sqrt{3}}( \vert 0, 2 \rangle + q^m
\vert 1 , 1 \rangle + q^{2m} \vert 2 , 0 \rangle).
\label{phasestates}
\end{equation}
It follows that the states $\vert m \rangle$ satisfy
\begin{equation}
E \vert m \rangle = e^{i\theta_m} \vert m \rangle \qquad \theta_m = \frac{2\pi m}{3},
\end{equation}
which shows that they are indeed phase states and $E$ is a unitary
phase operator. The phase states are orthonormal $(\langle m' \vert
m \rangle = \delta_{m',m})$. Then, Alice encodes her message in the
computational basis generating the Fock space ${\cal F}$ or in the
phase states basis $\{\vert m \rangle ~ , ~ m = 0, 1, 2\}$. It is
important to note that Alice can use two others bases which can be
generated from the phase states as  shown in what follows.
%%%%%%%%%%%%%%%%%%%%%%%%%%%%%%%%%%%%%%%%%%%%%%%%%%%%%%%%%%%%%%%%%%%%%%%%
 \subsection{Mutually unbiased bases }
 %%%%%%%%%%%%%%%%%%%%%%%%%%%%%%%%%%%%%%%%%%%%%%%%%%%%%%%%%%%%%%%%%%%%%%%

Recall that two $d$-dimensional bases are said to be unbiased if and
only if the modulus of the inner product of any vector of one basis
with any vector of the other one is equal to $1/{\sqrt d}$
\cite{Ivanovic}-\cite{WoottersFields}. The number $M$ of mutually
unbiased bases (MUBs) is such that $M \leq d+1$. The maximum number
$M = d+1$ can be achieved when $d$ is prime or a power of a prime
\cite{WoottersFields}. Construction of MUBs associated with finite
dimensional Hilbert space was considered via different methods as
for instance ones based on discrete Fourier analysis over Galois
fields and Galois rings, generalized Pauli matrices, angular
momentum theory, etc ( for a recent review and other methods of MUBs
construction see \cite{KibJPhysA0809}). Accordingly, the three level
optical system we are considering, has four mutually unbiased bases.
The first basis is the computational basis $B_3 := \{ \vert 2 , 0
\rangle ~ , ~ \vert 1 , 1 \rangle ~,~ \vert 0 , 2 \rangle \}$ and
the second one is  generated by the phase states
(\ref{phasestates}). Indeed, it is easy to check that the states
(\ref{phasestates}) are unbiased to the computational basis . The
two other remaining mutually unbiased bases can be generated as
follows. We introduce the time evolution operator $U(t)$ defined by
\begin{equation}
U(t)\vert n_{h} , n_{v} \rangle =  e^{ia^+_ha^-_h a^-_v a^+_v
t}\vert n_{h} , n_{v} \rangle
\end{equation}
in term of vertical  and horizontal ladder operators. The operator
$U(t)$  generates MUBs when acting on the phase states
(\ref{phasestates}). Indeed, for the discrete values
$$  t := t_p = \frac{p\pi}{3} ~~{\rm with} ~~p = 0, 1, 2,$$
the states (\ref{phasestates}) transform as
\begin{equation}
 U(t_p)\vert  m \rangle := \vert p , m \rangle =  \frac{1}{\sqrt{3}}( \vert 0, 2 \rangle
 + q^{m+p}
\vert 1 , 1 \rangle + q^{2m + p} \vert 2 , 0 \rangle).
\end{equation}
Notice that for $p = 0$, the states $\vert 0 , m \rangle $ coincide with
phase states (\ref{phasestates}) ($\vert 0 , m \rangle \equiv \vert  m \rangle$).
 It is simply verified that the computational basis $B_3$ and the
bases $B_p = \{ \vert p , m \rangle\}$ $( p = 0, 1, 2)$ are mutually
unbiased. To close this section, it is important to note that the
operator $U(t_p)$  is related to the quadratic discrete Fourier
transform \cite{KibJPhysA0809} and coincides with the so-called
tritter \cite{dotukowski} which is a generalization, in three
dimensional case, of the $2\times 2$ unitary operation
 characterizing a 50-50
beam splitter.

%%%%%%%%%%%%%%%%%%%%%%%%%%%%%%%%%%%%%%%%%%%%%%%%%%%%%%%%%%%%%%%%%%%%%%%%
 \section{ Eavesdropping strategy and mutual informations }
 %%%%%%%%%%%%%%%%%%%%%%%%%%%%%%%%%%%%%%%%%%%%%%%%%%%%%%%%%%%%%%%%%%%

As mentioned above,  quantum key distribution in the presence of
several eavesdroppers was developed in \cite{hamid}. This scenario
extends the BB84 protocol by investigating the effect of several
eavesdropper intercept-resend attacks on the quantum error and
mutual information between two legitimate parties. It was shown that
the secured-unsecured transition depends strongly on the number of
eavesdroppers and their probabilities of intercepting attacks. In
this section, we investigate the effect of several eavesdroppers on
the quantum cryptographic scheme using biphoton qutrits.

%%%%%%%%%%%%%%%%%%%%%%%%%%%%%%%%%%%%%%%%%%%%%%%%
\subsection{ Eavesdropping strategy: the Model}
%%%%%%%%%%%%%%%%%%%%%%%%%%%%%%%%%%%%%%%%%%%%%%

Following the idea discussed by Bourenanne et al \cite{Bourennane},
we consider a protocol with $M$ $(2 \leq M \leq 4)$ mutually
complementary bases and $3$ orthogonal states in each base.  We
shall assume that this protocol is  under attack by an arbitrary
number $N$ of eavesdroppers $E_1$ , $E_2$ , ..., $E_N$. Within  this
protocol, Alice first selects  randomly one of the $M$ bases in
which she wants to encode her state and second decides which of the
$3$ optical states ($\vert 0,2 \rangle, \vert 1,1 \rangle ~{\rm or}~
\vert 2, 0\rangle$) to send.  In other words, she sends to Bob
random states in which the number of horizontally polarized photons
is $0, 1, 2$ with equal probability of $1/3$. Bob measures each
symbol sent by selecting at random between the $M$ bases. Hence, the
mutual information between Alice and Bob can be described by a joint
probability
 $P( x_A , x_B )$. The random variables $x_A = 0 , 1, 2$ and $x_B = 0 , 1, 2$
 represent the number of horizontally polarized photons
prepared by the sender (Alice) and the measurement results obtained
by the receiver (Bob). Between them a number $N$ of eavesdropper
$E_i$  $(i = 1,..., N )$ are trying to intercept the sent message.
Each eavesdropper $E_i$ intercepts, with probability $\omega_i$, the
biphoton state emitted by the eavesdropper $E_{i-1}$. He or she
measures its number of photons horizontally polarized by selecting
at random, with probability $1/M$, between the $M$ MUBs and resends
it, in its measured state, to the eavesdropper $E_{i+1}$. At the
place of the non measured biphoton polarization, with probability $1
- \omega_i$, the eavesdropper $E_i$ sends randomly $0, 1, 2$, with
equal probability $1/3$, to the eavesdropper $E_{i+1}$. In the same
way, the eavesdropper $E_{i+1}$ intercepts, with probability
$\omega_{i+1}$, the biphoton state emitted by the eavesdropper
$E_i$, measures its number of photons polarized horizontally by
selecting at random, with probability $1/M$, between the $M$ MUBs,
and resends it in its measured state to the eavesdropper $E_{i + 2}$
and so on. We note that the eavesdropper $E_1$ intercepts the state
emitted by Alice and the eavesdropper $E_N$ resends the biphoton to
Bob.

Finally, in order to obtain a secret key, Alice and Bob use an
authenticated public channel to estimate the error rate and the
maximal quantity of information obtained by the eavesdroppers.
However, if the error rate (called the error probability) is greater
than a critical value (quantum error) Alice and Bob begin  another
protocol to establish another secret key until the error rate
becomes smaller than the quantum error.

 %%%%%%%%%%%%%%%%%%%%%%%%%%%%%%%%%%%%%%%%%%%%%%%%%%%%%%%%%%%%%%%%%%%%%%%%
 \subsection{The mutual informations}
 %%%%%%%%%%%%%%%%%%%%%%%%%%%%%%%%%%%%%%%%%%%%%%%%%%%%%%%%%%%%%%%%%%%

To evaluate the mutual information between Alice and Bob and the
amount of information gained by the eavesdroppers, the relevant
information is the Shannon information of the sifted symbols, i.e.,
the symbols for which Alice
and Bob have used the same bases. This information is measured in bits for simplicity.\\
Let us denote by $p(x)$  the prior probability for Alice to send the
symbol $x$ and $p(x\vert y)$ is the posteriori probability that is
the conditional probability of the sending party  having sent the
symbol $x$ and the receiver (Bob or Eves) measured the result $y$.
The mutual information is (see for instance \cite{Bourennane})
\begin{equation}
I_{AY} =  \log_2 3 - H_{\rm apost}
\end{equation}
where $Y$ stands for $B, E_1, E_2, \cdots, E_N$ and the quantity
\begin{equation}
H_{\rm apost} = - \sum_y p(y) \sum_x p(x\vert y) \log_2 p(x\vert y)
\end{equation}
is the aposteriori entropy.  Using the symmetry properties of the
protocol, it is easy to check that
\begin{equation}
I_{AY} =  \log_2 3 + \sum_{n_h=0}^{2} p(n_h\vert 0) \log_2 p(n_h\vert
0).
\label{IAY}
\end{equation}
The mutual information between Alice and Bob $I_{AB}$ and between
Alice and the $m$-th eavesdropper $I_{AE_m}$
 $ (m = 1, 2, \cdots, N)$ are expressed in terms of the conditional probabilities which can be easily evaluated  in terms
of the attack probabilities using the eavesdropping strategy
discussed above.

Using the expression (\ref{IAY}), we can hence obtain the mutual information between Alice and Bob
\begin{equation}
I_{AB} = \log_2 3 + P_{AB} (0 \vert 0) \log_2 ( P_{AB} (0 \vert
0)) + [ 1- P_{AB} (0 \vert 0)] \log_2 \bigg[\frac{ 1- P_{AB} (0
\vert 0)}{2}\bigg],
\label{IAB}
\end{equation}
in term of the conditional probability $P_{AB} (0 \vert 0)$. Based on the assumptions defining the eavesdropping strategy, one can
show that this probability takes  the following form
\begin{equation}
 P_{AB} (0 \vert 0) = \sum_{k=0}^{N} a_k(N) \Omega_k(N)
 \label{PAB}
\end{equation}
where the coefficients $a_k(N)$ are given by
\begin{equation}
  a_k(N) = \frac{1}{M^{N-k}}\bigg[ 1 + \frac{M-1}{3}\sum_{i=0}^{N-k-1}M^i\bigg],
\end{equation}
and the quantities $\Omega_k(N)$ are expressed in term of eavesdroppers attack probabilities as
\begin{equation}
  \Omega_k(N) = \omega_1\omega_2\cdots\omega_N \sum_{i_1<i_2<\cdots<i_k}\bigg[ \frac{1-\omega_{i_1}}{\omega_{i_1}}
\frac{1-\omega_{i_2}}{\omega_{i_2}}\cdots\frac{1-\omega_{i_k}}{\omega_{i_k}}\bigg]
\end{equation}
for $k\neq 0$ and  the indices $i_j$ take the values $ 1, \cdots ,
N$. For $ k = 0$,
\begin{equation}
 \Omega_0 (N) = \omega_1\omega_2\cdots\omega_N.
\end{equation}

 Similarly, using equation  (\ref{IAY}), it is simple to show that the direct reconciliation information between Alice and
the $m$-th eavesdropper is given by
\begin{equation}
I_{AE_m} = \log_2 3 + P_{AE_m} (0 \vert 0) \log_2 ( P_{AE_m} (0
\vert 0)) + [ 1- P_{AE_m} (0 \vert 0)] \log_2 \bigg[\frac{ 1-
P_{AE_m} (0 \vert 0)}{2}\bigg], \label{IAEm}
\end{equation}
where
\begin{equation}
 P_{AE_m} (0 \vert 0) =  \frac{1-\omega_m}{3} + \sum_{k=0}^{m-1} a_k(m)
 \Omega_k(m).
  \label{PAEm}
\end{equation}

The lost information between the honest parties Alice and Bob
corresponds to the maximum information intercepted by the entire
eavesdroppers. This is given by
\begin{equation}
I_{AE} = {\rm Max}\bigg(I_{AE_1}, I_{AE_2},\cdots,I_{AE_{N-1}} ,I_{AE_N}\bigg).
\label{IAEmax}
\end{equation}
The error rate or the error probability ${\rm P_{err}}$ is defined by
\begin{equation}
{\rm P_{err}}= \sum_{x_A , x_B} \bigg{\vert}
P_{AB}(x_A/x_B)\big{\vert}_{\omega_i = 0} -
P_{AB}(x_A/x_B)\big{\vert}_{\omega_i \neq 0}\bigg{\vert}.
\end{equation}
The quantum error ${\rm Q_{err}}$ is the value of the error
probability ${\rm P_{err}}$ for which $ I_{AB} = I_{AE}$ .
It follows that  for ${\rm P_{err}} < {\rm Q_{err}}$  we have $I_{AE} < I_{AB} $, while for ${\rm P_{err}} > {\rm Q_{err}}$  we have $I_{AE}
> I_{AB}$.

 %%%%%%%%%%%%%%%%%%%%%%%%%%%%%%%%%%%%%%%%%%%%%%%%%%%%%%%%%%%%%%%%%%%%%%%%
 \section{Results and discussion}
 %%%%%%%%%%%%%%%%%%%%%%%%%%%%%%%%%%%%%%%%%%%%%%%%%%%%%%%%%%%%%%%%%%%
As illustration of the analysis developed in the previous section,
we investigate in what follows the security of  the protocol based
on optical qutrit states in the presence of several eavesdroppers,
in some particular cases.
%%%%%%%%%%%%%%%%%%%%%%%%%%%%%%%%%%%%%%%%%%%%%%%%%%%%%%%%%%%%%%%%%%%%%%%%
 \subsection{One eavesdropper and mutual informations}
 %%%%%%%%%%%%%%%%%%%%%%%%%%%%%%%%%%%%%%%%%%%%%%%%%%%%%%%%%%%%%%%%%%%
 %%%%%%%%%%%%%%%%%%%%%%%%%%%%%%%%%%%%%%%%%%%%%%%%%%%%%%%%%%%%%%%%%%%%%%%%
In the situation where only one eavesdropper is trying to intercept
the message sent by Alice ($\omega_1 = \omega$), the equation
(\ref{PAB}) gives
\begin{equation}
 P_{AB} (0 \vert 0) = \frac{1}{M}\bigg(1+\frac{M-1}{3}\bigg)\omega +
 (1-\omega).
\label{pab}
\end{equation}
Similarly from equation (\ref{PAEm}), one obtains
\begin{equation}
 P_{AE} (0 \vert 0) =  \frac{1-\omega}{3} +  \frac{1}{M}\bigg(1+\frac{M-1}{3}\bigg)\omega.
\label{pae}
\end{equation}

Reporting the conditional probabilities (\ref{pab}) and (\ref{pae})
in the equations (\ref{IAB}) and (\ref{IAEm}), respectively, one has
the mutual information $I_{AB}$ and $I_{AE}$. The behavior of
$I_{AB}$ and $I_{AE}$ as function of the attack probability $\omega$
are plotted in the figure 1.a.
\begin{center}
  \includegraphics[width=3in]{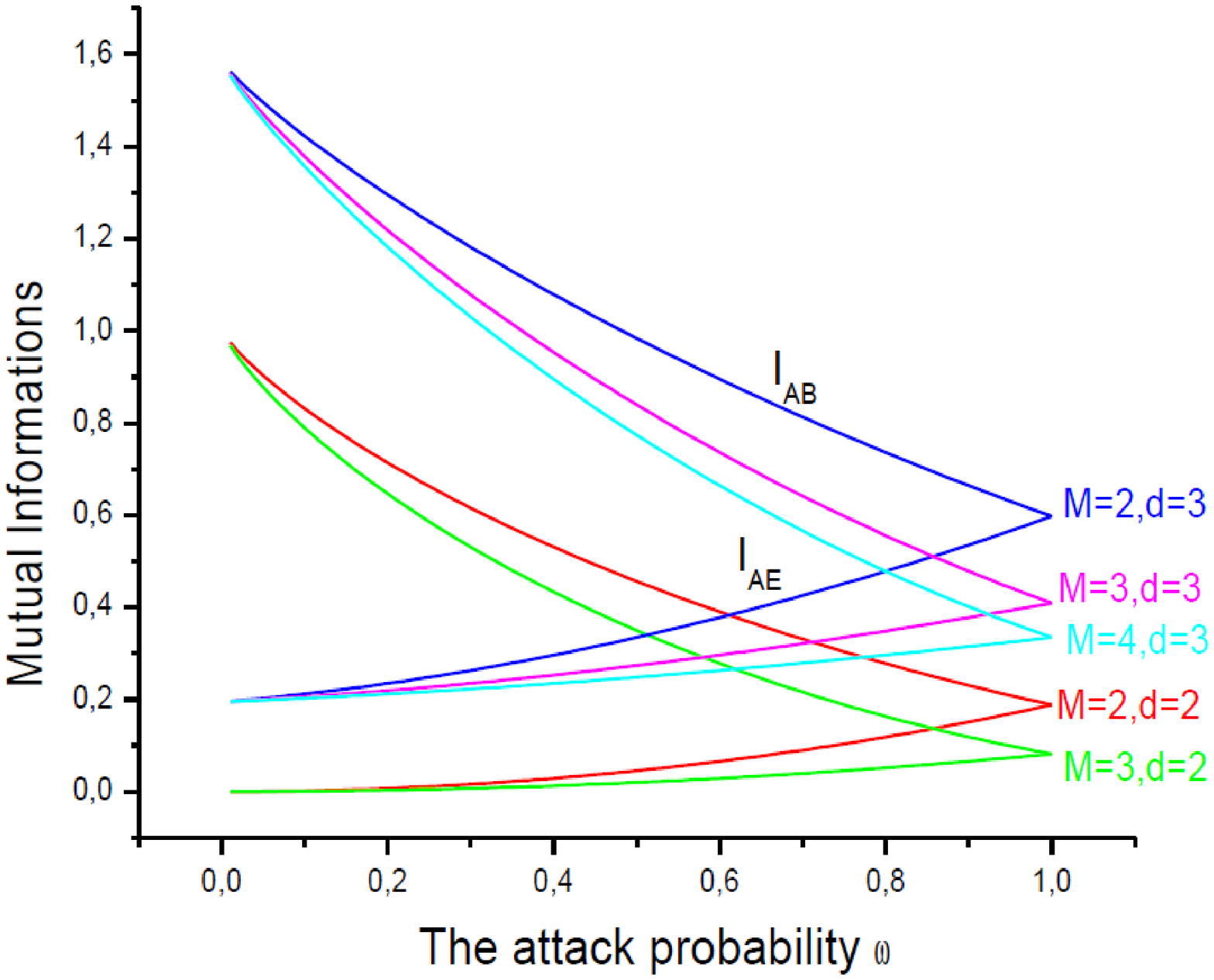}\\
FIG. 1.a:  {\sf Mutual information $I_{AB}$ and $I_{AE}$ as a
function of the attack probability $\omega$ for qubits and qutrits.}
\end{center}

As expected, the mutual information between Alice and Bob and the
amount of information intercepted by Eve coincide for $\omega = 1$.
In figure 1.a, we also plotted the mutual information $I_{AB}$ and
$I_{AE}$ when the sender uses a two dimensional system with two or
three mutually unbiased bases to encode her message. The explicit
expressions of  $I_{AB}$ and $I_{AE}$  are given in the appendix.
 They are computed similarly to ones derived in the
previous section for a three level system.   This helps us to
compare the amount of mutual information when Alice uses qubits or
qutrits as it shown in the figure 1.a. Note that the obtained mutual
information for $d=2$ and $M=2$ are in agreement with the results
derived in
\cite{hamid}.\\

To examine the security of the protocol, we studied the  mutual
information between the legitimate parties Alice and Bob $I_{AB}$
and lost information $I_{AE}$ as function of the error probability
(Figure 1.b).
\begin{center}
  \includegraphics[width=3.5in]{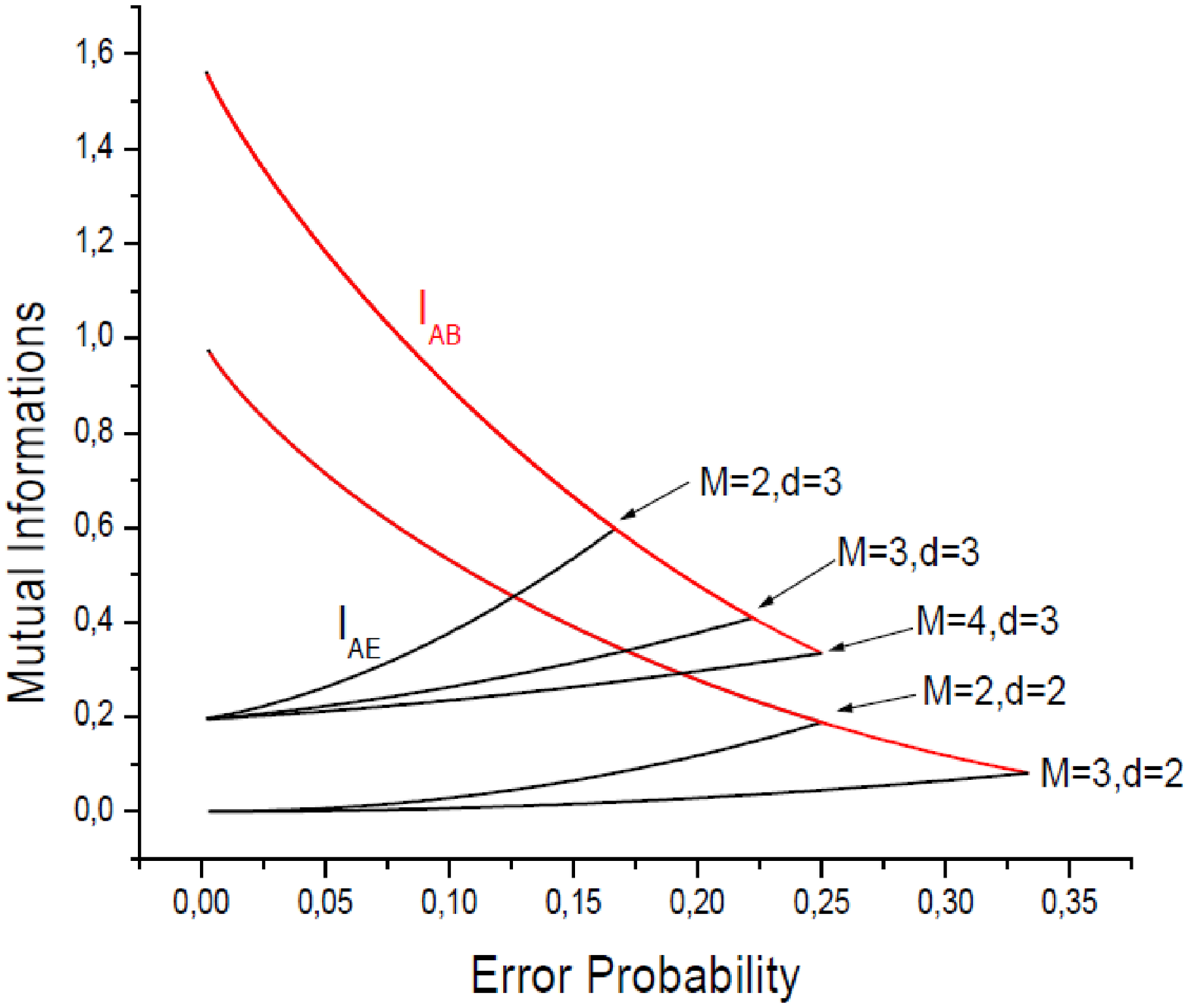}\\
FIG. 1.b:  {\sf Mutual information $I_{AB}$ and $I_{AE}$ as a
function of quantum error ${\rm Q_{err}}$ for qubits and qutrits.}
\end{center}

In this figure, we give the behavior of $I_{AB}$ and $I_{AE}$  for
$d=3$ when Alice chooses to encode her message using two, three or
four complementary bases in the presence of only one eavesdropper.
We evaluate
 the quantum error for each case. The results are summarized in table 1.
In figure 1.b, we also present the information $I_{AB}$ and $I_{AE}$
for qubits $(d = 2)$ as a function of error probability. This is
useful in order to compare our results with ones obtained within the
BB84 protocol and the protocol involving qubits with six states
$(M=3)$ ( see \cite{Bruss PRL 98} and \cite{Bechmann1}). In this
particular case, we evaluate the quantum error for $M=2$ and $M=3$.
The results are given in table 1. It is clear from the figure as
well as the table 1 that the protocol using $(d = 2)$ with $(M=3)$
appears to provide better security. It is also
 important to stress that the quantum error in the case $d=3$ and $M=4$ provides the
 same quantum error as in the BB84 protocol.  This shows that in a protocol involving three
level systems, Alice should use four mutually unbiased bases to
ensure the security of her sent information. It must be noticed that
for $d$ fixed, the quantum error increases with increasing values of
the number of mutually unbiased bases $M$. In this respect, Alice
must encode her message using all the available mutually unbiased
bases to minimize the eavesdropping effects.\\ \\

\begin{center}
\begin{tabular}{|c|c|c|c|c|}
   \hline
   %~~
   $M$ &  $2$ & $3$ & $4$  \\
   \hline %\hline
   $ d = 3 $ & $0.167$ & $0.222$ &
$0.250$   \\
   \hline
 $ d = 2 $ & $0.25$ & $0.335$  &
  \\
   \hline
\end{tabular}\\
{\vspace {0.5cm}} {\sf Table 1: Quantum error for two and three
level systems.}
\end{center}

%%%%%%%%%%%%%%%%%%%%%%%%%%%%%%%%%%%%%%%%%%%%%%%%%%%%%%%%%%%%%%%%%%%%%%%%
 \subsection{Two eavesdroppers and mutual information}
 %%%%%%%%%%%%%%%%%%%%%%%%%%%%%%%%%%%%%%%%%%%%%%%%%%%%%%%%%%%%%%%%%%%

Now, we consider the situation where two eavesdroppers $E_1$ and
$E_2$ attack with probabilities $\omega_1$ and $\omega_2$,
respectively. In this case, the equation (\ref{PAB}) takes the
simple form
\begin{equation}
 P_{AB} (0 \vert 0) = \frac{1}{M^2}\bigg(1+ \frac{M^2 -
 1}{3}\bigg)\omega_1\omega_2 + \frac{1}{M}\bigg(1+ \frac{M -
 1}{3}\bigg)\bigg(\omega_1(1-\omega_2) + (1-\omega_1)\omega_2\bigg)+
 (1-\omega_1)(1-\omega_2).
\label{pab2}
\end{equation}
Similarly, from  equations (\ref{PAEm}), one has
\begin{equation}
 P_{AE_1} (0 \vert 0) =  \frac{1-\omega_1}{3} + \frac{1}{M^2}\bigg(1+ \frac{M^2 -
 1}{3}\bigg)\omega_1\omega_2,
\label{pae1}
\end{equation}
and
\begin{equation}
 P_{AE_2} (0 \vert 0) =  \frac{1-\omega_2}{3} + \frac{1}{M^2}\bigg(1+ \frac{M^2 -
 1}{3}\bigg)\omega_1\omega_2 + \frac{1}{M}\bigg(1+ \frac{M -
 1}{3}\bigg)\bigg( (1-\omega_1)\omega_2\bigg).
\label{pae2}
\end{equation}
Substituting (\ref{pab2}) into (\ref{IAB}), one gets the mutual
information $I_{AB}$ between Alice and Bob. The conditional
probabilities (\ref{pae1}) and (\ref{pae2}) together with the
equations (\ref{IAEm}) and (\ref{IAEmax}) give the lost information
to the eavesdropper, $I_{AE}$. The figure 2.a  represents the mutual
information $I_{AB}$ and $ I_{AE}$ (for $d = 3$ and $M=2$) as
function of the attack probability $\omega_1$ of the first
eavesdropper for different values of the attack probability of the
second eavesdropper.
\begin{center}
  \includegraphics[width=3in]{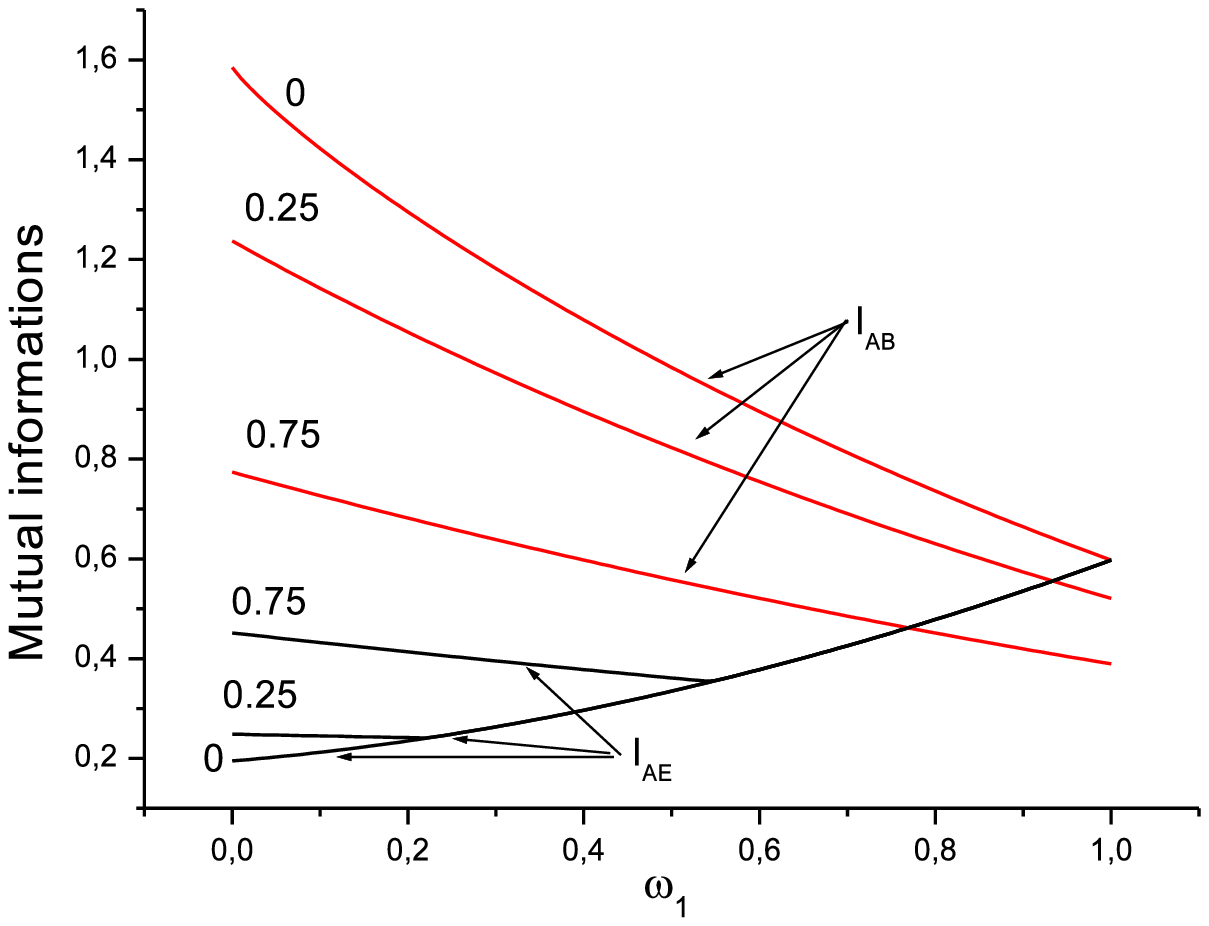}\\
FIG. 2.a:  {\sf Mutual information $I_{AB}$ and $I_{AE}$ as a
function of the attack probability $\omega_1$ for different values
of $\omega_2$.}
\end{center}
It clear that $I_{AB}$ and $I_{AE}$ depend strongly on the values
$\omega_1$ and $\omega_2$ and the mutual information between the
legitimate parties decreases as $\omega_1$ and $\omega_2$ increase.
The limiting case $\omega_2 = 0$ corresponds to the protocol
involving only one eavesdropper. In this case, the amount of
information gained by the eavesdropper increases to intercept the
information exchanged by Alice and Bob at $\omega_1 = 1$. But, when
$\omega_2 \neq 0$, the information intercepted by the eavesdroppers
intersect $I_{AB}$ for $\omega_{1{\rm tr}} \simeq 0.92$ when
$\omega_2 = 0.25$ and for  $\omega_{1{\rm tr}} \simeq 0.76$ when
$\omega_2 = 0.75$. Hence, for $\omega_1 > \omega_{1{\rm tr}}$, the
amount of information lost becomes greater that one exchanged
between Alice and Bob and occurs at the transition from  the secured
to unsecured phase.

To understand the security of quantum key distribution based on
qutrits in presence of two eavesdroppers, we studied the transition
between the secured and unsecured phases. We note that in the
secured phase, the error probability is smaller than the quantum
error, while in the no secured phase the error probability is
greater than the quantum error. At the transition line, the error
probability coincides with the quantum error. Phase diagram, in the
space parameter $(\omega_1 , \omega_2)$, is presented in the figure
2.b. This  shows the transition line between secured and no secured
phases. In contrast to the case of the protocol with one
eavesdropper for which the secured-unsecured transition occurs at an
intercept probability $\omega_1 = 1$, the region of secured phase
depends on both intercept probability rates $\omega_1$ and
$\omega_2$. We consider the situations where Alice uses two, three
and four mutually bases.

From figure 2.b, it is easily seen  that
 for $0<\omega_1 < 0.55$, the line
 transition between the secured and unsecured phase is
 $\omega_2$-independent. It is also independent of the number
 of mutually independent bases $M$ used by Alice. This changes
 drastically when the probability attack of the eavesdropper $E_1$ becomes
grater than $0.55$. In this case, the transition depends strongly on
the values of attacks probability of the second eavesdropper as well
as the mutually unbiased bases $M$.  The transition probability, at
the transition line, increases with decreasing $ \omega_2$
\begin{center}
  \includegraphics[width=3in]{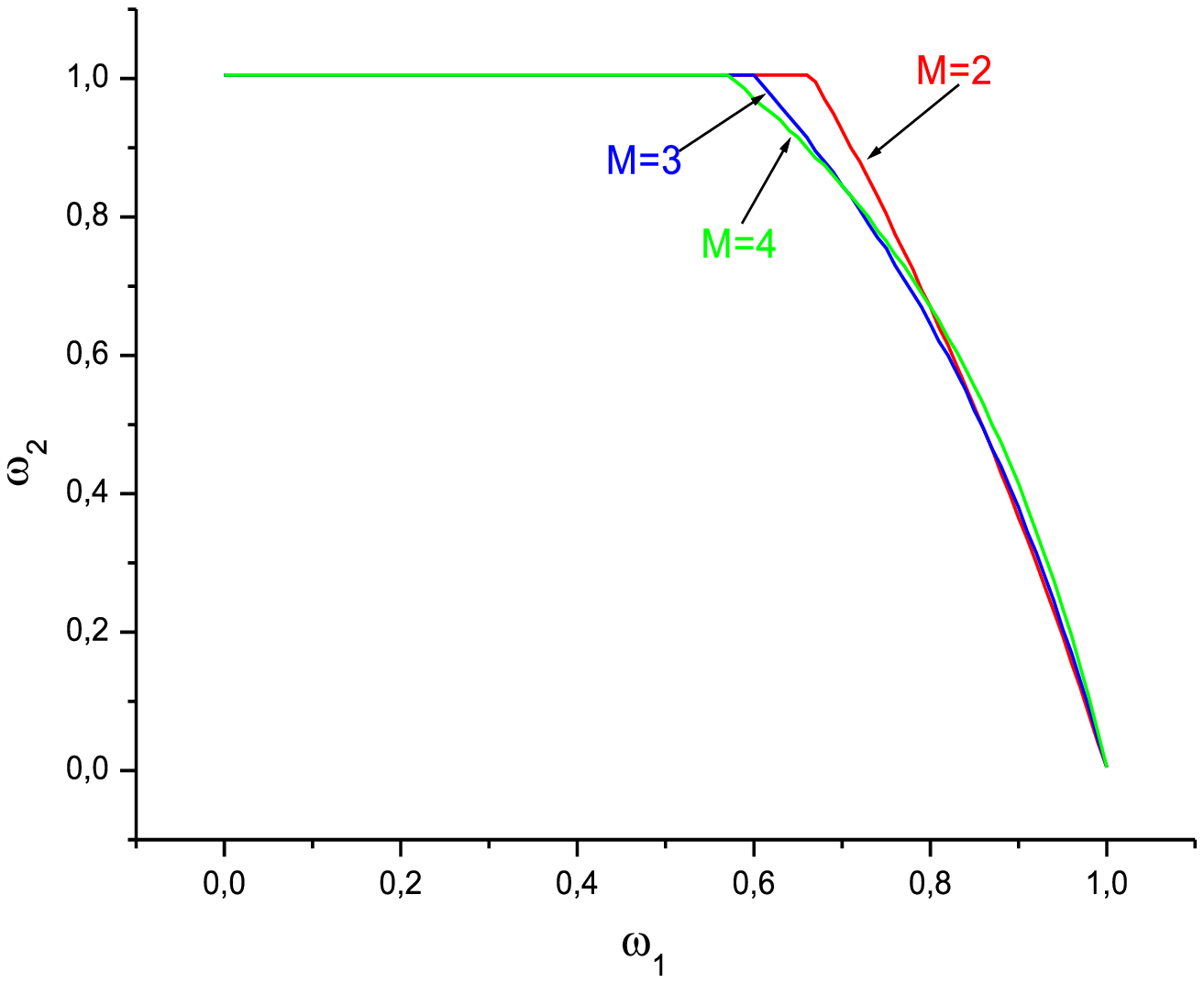}\\
FIG. 2.b:  {\sf The $(\omega_1 , \omega_2)$ phase diagram for
qutrits.}
\end{center}

To clarify the behavior of the transition from the secured to
unsecured phase for  $0.55 \leq \omega_1 \leq 1$, we give the figure
2.c where the line transition is represented for $M = 2, M= 3$ and
$M = 4$. From this figure, one can see that for $0.55 <\omega_1 <
0.67$, the protocol involving $M = 4$ complementary bases gives less
security than the two others using two and three bases. This
situation becomes different in the region $0.67 <\omega_1 < 0.86$;
Indeed, in this case the model with $M=3$ provides less security.
Finally, for $\omega_1 > 0.86$, when Alice uses only two
complementary bases, the area of secured phase is reduced in
comparison to ones corresponding to the secured phases obtained with
$M = 3$ and $M = 4$.
\begin{center}
  \includegraphics[width=3in]{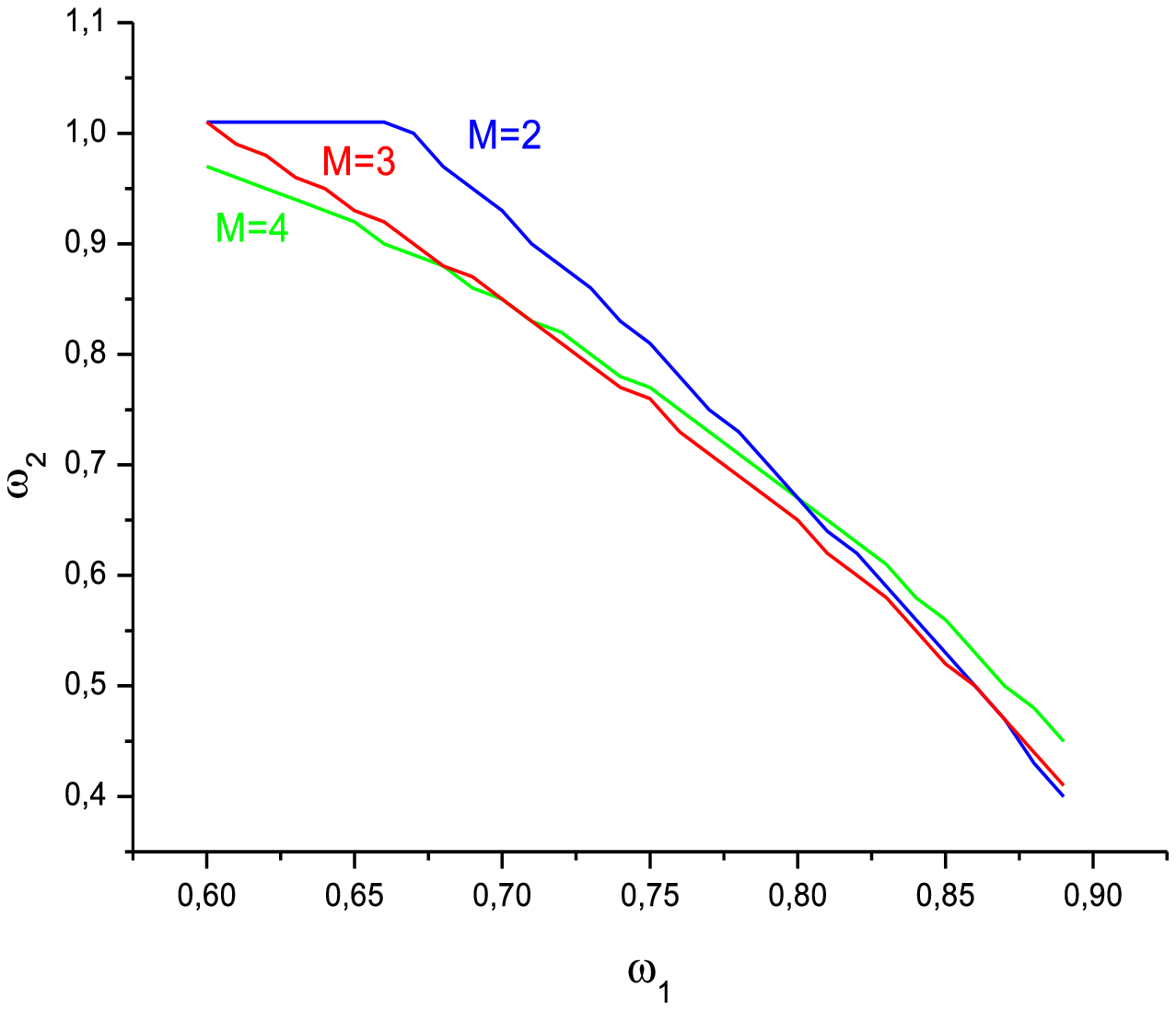}\\
FIG. 2.c:  {\sf The $(\omega_1 , \omega_2)$ phase diagram for $0.55
\leq \omega_1 \leq 1$.}
\end{center}
%%%%%%%%%%%%%%%%%%%%%%%%%%%%%%%%%%%%%%%%%%%%%%%%%%%%%%%%%%%%%%%%%%%%%%%%
 \subsection{Many eavesdroppers }
 %%%%%%%%%%%%%%%%%%%%%%%%%%%%%%%%%%%%%%%%%%%%%%%%%%%%%%%%%%%%%%%%%%%

Now, we shall discuss the situation where the eavesdroppers
communicate between
 them and try to intercept the same state with identical probability
 $\omega_i = \omega $ for $ i = 1, 2, \cdots , N$. In this case, the equation
 (\ref{PAB}) takes the simple form
\begin{equation}
P_{AB}(0/0) = \frac{1}{3} + \frac{2}{3}\bigg[ 1 + \frac{\omega}{M} (
1 - M)\bigg]^N,
\end{equation}
and the equation (\ref{PAEm}) gives
\begin{equation}
P_{AE_m}(0/0) = \frac{1}{3} + \frac{2}{3}\frac{\omega}{M}\bigg[ 1 +
\frac{\omega}{M} ( 1 - M)\bigg]^{m-1}.
\end{equation}

The $(\omega, N)$-phase diagram (Figure 3.a) shows the
secured-unsecured transition when the $N$ eavesdroppers are
collaborating and attack with the same probability $\omega$.
\begin{center}
  \includegraphics[width=3in]{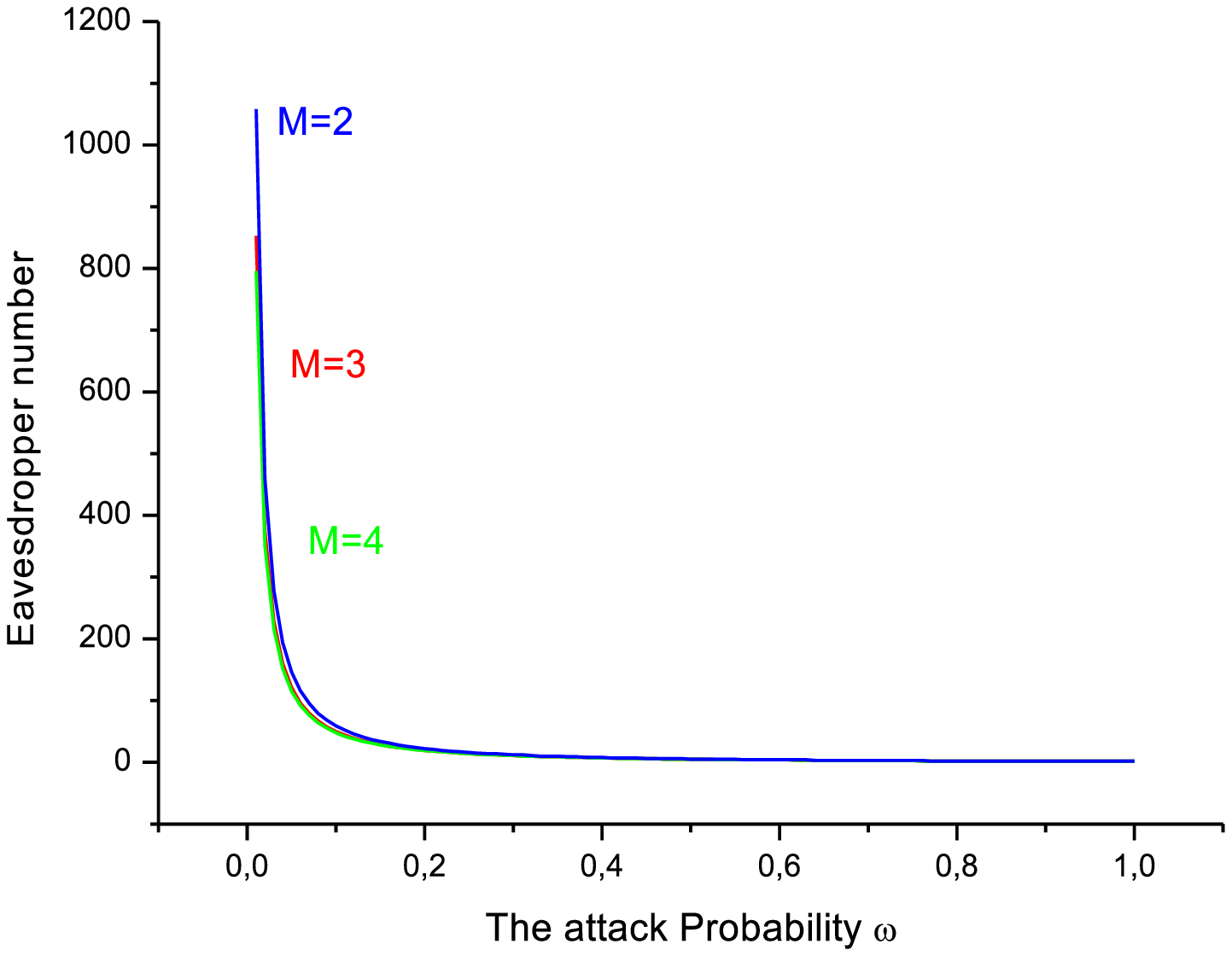}\\
FIG. 3a:  {\sf The $(\omega, N)$-phase diagram.}
\end{center}

The security of the protocol decreases when $N$ increases. Moreover,
in the region
 $(0.1 \leq \omega \leq 1, 0 \leq N \leq 50)$, the security of the protocol is completely
independent of the number $M$ of mutually unbiased bases used by the
sender. However, for $ N \geq 50$ and $0 \leq \omega \leq 0.1$ (see
figure 3.b), more security is provided by the protocol involving two
complementary bases (i.e., $M = 2$).
\begin{center}
  \includegraphics[width=3in]{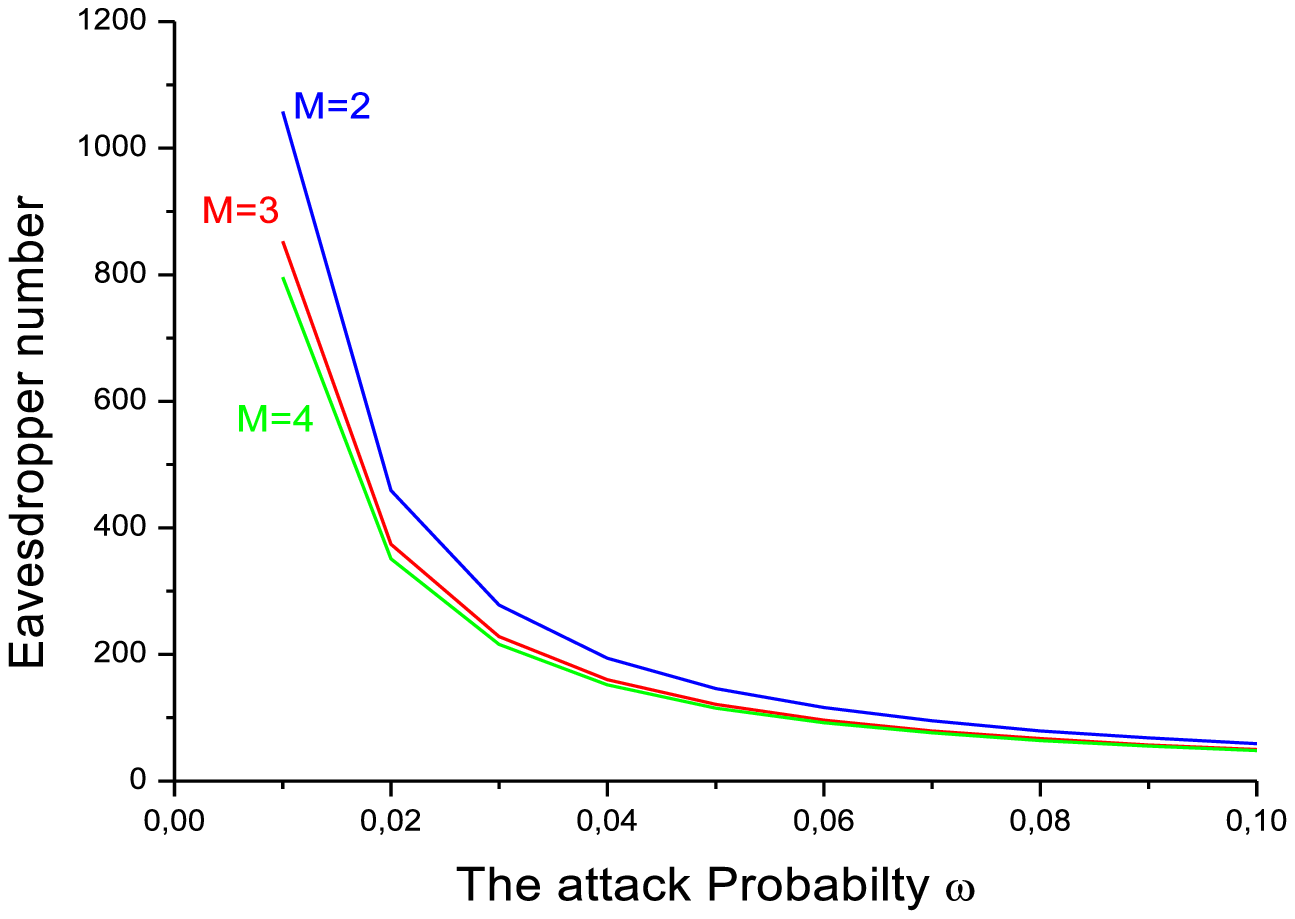}\\
FIG. 3b:  {\sf The $(\omega, N)$-phase diagram.}
\end{center}

 %%%%%%%%%%%%%%%%%%%%%%%%%%%%%%%%%%%%%%%%%%%%%%%%%%%%%%%%%%%%%%%%%%%%%%%%
 \section{Concluding remarks}
 %%%%%%%%%%%%%%%%%%%%%%%%%%%%%%%%%%%%%%%%%%%%%%%%%%%%%%%%%%%%%%%%%%%
We have considered quantum cryptographic scheme where bi-photon
qutrits are used to encode the information. Using the Pegg-Barnett
phase operator, we defined the four mutually unbiased bases of this
three level system. This is mainly based on the phase states
approach. To compare the qutrits based cryptographic protocols with
their qubits based counterpart, we also investigated the
bi-dimensional quantum systems when the sender uses two or three
bases  in the presence of many eavesdroppers. In the situation where
only one eavesdropper is involved, for a two level quantum system,
the protocol based on three complementary bases is safer than the
BB84 one which uses two bases only. We have shown that when Alice
and Bob exchange information using qutrits, the safer scenario
corresponds to one using four mutually unbiased bases. It follows
that for $d =2$ as well as $d =3$, to ensure the security of the
sent information, Alice should encode her message using all the
available complementary bases of the Hilbert space of the quantum
system under consideration. It is important to note that the quantum
error for the scheme involving four mutually unbiased bases
coincides with the one obtained in BB84 protocol. It seems that for
higher dimensional quantum systems, the maximal security is reached
when the sender uses all available complementary bases. These
results change drastically when more than one eavesdropper attempt
to intercept the information exchanged between the legitimate
parties. Indeed, the secured-unsecured transition for qutrits is
strongly dependent on the attack probabilities $\omega_1$ and
$\omega_2$. In this case many scenarios are possible and there is a
concurrence between the protocols involving two, three or four
mutually unbiased bases.
 Finally, we have shown that the number  of the eavesdroppers reduces
 the area of secured phase and the protocol becomes less
secure. In particular, we examined the case where a large number of
collaborating eavesdroppers are trying to intercept the information
with equal probability $\omega$. This shows clearly that the number
of eavesdroppers is very important in dealing with the security of
any quantum cryptographic key distribution protocol .

{\vskip 1.5cm}
%%%%%%%%%%%%%%%%%%%%%%%%%%%%%%%%%
\noindent {\bf Acknowledgments}\\
%%%%%%%%%%%%%%%%%%%%%%%%%%%%%%%%%%%
MD  would like to thank the hospitality and kindness extended to
him by the Max Planck Institute for Physics of Complex Systems (Dresden, Germany)
where this work was done.

{\vskip 1.5cm}

%%%%%%%%%%%%%%%%%%%%%%%%%%%%%%%
\noindent {\bf Appendix}\\
%%%%%%%%%%%%%%%%%%%%%%%%%%%%%%%%

It is well known that for a two dimensional quantum system, there is three maximally unbiased bases from
 which Alice can choose $M = 2$ or $M = 3$ bases to encode her message. Similarly to the qutrits case
discussed above and using the same assumptions, the mutual information between Alice and Bob, in presence of one
eavesdropper only, can be found as

$$ I_{AB} = 1 + P_{AB} (0 \vert 0) \log_2 ( P_{AB} (0 \vert
0)) + [ 1- P_{AB} (0 \vert 0)] \log_2 \bigg[\frac{ 1- P_{AB} (0
\vert 0)}{2}\bigg],$$
where
$$ P_{AB}(0\vert 0) =  1 - \frac{\omega}{2}\frac{M-1}{M}.$$
% \frac{1}{M}\bigg(1+\frac{M-1}{2}\bigg)\omega + (1-\omega)
The intercepted information by the eavesdropper is

$$I_{AE}(0\vert 0) = 1 + P_{AE} (0 \vert 0) \log_2 ( P_{AE} (0
\vert 0)) + [ 1- P_{AE} (0 \vert 0)] \log_2 \bigg[\frac{ 1- P_{AE}
(0 \vert 0)}{2}\bigg],$$
where
$$P_{AE} =  \frac{1}{2}\bigg( 1 + \frac{\omega}{M}\bigg).$$
%\frac{1-\omega}{2} +  \frac{1}{M}\bigg(1+\frac{M-1}{2}\bigg)\omega
For two eavesdroppers $E_1$ and $E_2$, trying to intercept the sent
information with probabilities $\omega_1$ and $\omega_2$, the
conditional probabilities are given by

$$P_{AB} (0\vert 0) = 1 - \frac{M-1}{2M}(\omega_1 + \omega_2) + \frac{(M-1)^2}{2M^2}\omega_1\omega_2, $$

%$$P_{AB} (0\vert 0) = \frac{1}{M^2}\bigg(1+ \frac{M^2 -
% 1}{2}\bigg)\omega_1\omega_2 + \frac{1}{M}\bigg(1+ \frac{M -
% 1}{2}\bigg)\bigg(\omega_1(1-\omega_2) + (1-\omega_1)\omega_2\bigg)+
% (1-\omega_1)(1-\omega_2),$$

$$P_{AE_1} (0\vert 0) =   \frac{1}{2}\bigg[ 1 - \omega_1(1-\omega_2) +\frac{\omega_1\omega_2}{M^2} \bigg],$$

%$$P_{AE_1} (0\vert 0) =   \frac{1-\omega_1}{2} + \frac{1}{M^2}\bigg(1+ \frac{M^2 -
% 1}{2}\bigg)\omega_1\omega_2$$

and

$$P_{AE_2} (0\vert 0) =  \frac{1}{2}\bigg[ 1 + \frac{\omega_2}{M} - \frac{M-1}{M^2}\omega_1\omega_2\bigg]$$

%$$P_{AE_2} (0\vert 0) =  \frac{1-\omega_2}{2} + \frac{1}{M^2}\bigg(1+ \frac{M^2 -
% 1}{2}\bigg)\omega_1\omega_2 + \frac{1}{M}\bigg(1+ \frac{M -
% 1}{2}\bigg)\bigg( (1-\omega_1)\omega_2\bigg)$$

from which one can evaluate the mutual informations $I_{AB}$, $I_{AE_1}$ and $I_{AE_2}$.

\end{document}